\documentclass[aps,floatfix]{revtex4}
\usepackage{color}
\usepackage{graphicx}
\usepackage{epsfig}
\usepackage{dcolumn}
\usepackage{bm}
\usepackage{mathtools}
\usepackage{amssymb}
\usepackage{calrsfs}
\DeclareMathAlphabet{\pazocal}{OMS}{zplm}{m}{n} 
\usepackage{amsmath}
\usepackage{subfigure}
\newcommand{\be}{\begin{equation}}
\newcommand{\ee}{\end{equation}}
\newcommand{\bea}{\begin{eqnarray}}
\newcommand{\eea}{\end{eqnarray}}
\newcommand{\rt}{{\vec r}}
\newcommand{\astW}{\stackrel{\ast}{W}}

\newcommand{\tK}{{\tilde K}}
\newcommand{\tQ}{{\tilde Q}}
\newcommand{\tS}{{\tilde S}}

\newcommand{\pro}{\partial}

\newcommand{\ba}{\begin{array}}
\newcommand{\ea}{\end{array}}

\newcommand{\nn}{\nonumber}

\newcommand{\Ka}{\mathcal{K}}

\newcommand{\Da}{\mathcal{D}}

\newcommand{\Fa}{\mathcal{F}}

\begin{document}
\title{Axially-symmetric stationary solutions in a pure $SU(3)$ QCD}
\bigskip
\author{D. G. Pak}
\affiliation{Chern Institute of Mathematics, Nankai University,
Tianjin 300071, China}
\author{P.M. Zhang}
\affiliation{Institute of Modern Physics, Chinese Academy of Sciences,
Lanzhou 730000, China}
\begin{abstract}
We propose an ansatz for a class of regular axially-symmetric solutions
in $SU(3)$ QCD. After averaging over time period the solution can be treated as a
non-topological monopole-antimonopole pair. We demonstrate that QCD Lagrangian 
on the space of such solutions is explicitly Weyl symmetric and reduces to a generalized
$\phi^4$ model with four independent fields.  All solutions possess quantum stability
under vacuum gluon fluctuations. 
\end{abstract}
\vspace{0.3cm}
\maketitle

\section{Introduction}

Generation of a non-trivial vacuum due to color monopole condensation
in a dual superconductor is one of the most appealing mechanisms of confinement 
in quantum chromodynamics (QCD) \cite{nambu74,mandelstam76,polyakov77,thooft81}.
The first attempt to realize such a scenario had been undertaken in the Savvidy vacuum 
\cite{savv} based on homogeneous chromomagnetic vacuum field. It was shown that the 
vacuum is unstable due to presence of a tachyonic unstable mode \cite{N-O}.
In subsequent studies several vacuum models have been proposed with various implemented
vacuum field configurations: the vortices \cite{niel-nino,niel-oles,amb-oles1,amb-oles2,chernodub14}, 
center vortices \cite{centervort1,centervort2,centervort3}, monopoles \cite{cho2010,cho2014,chernodub05},
dyons  \cite{diak-petrov} etc. 
Recently it has been proposed that regular stationary spherically symmetric monopole and
axially symmetric monopole-like solutions are stable under the vacuum gluon fluctuations
at microscopic space-time scale \cite{ijmpa2017,plb2017}.  This gives a hope that
such solutions  can serve as structure elements in constructing the true QCD vacuum. 
 
In the present paper we describe a general class of regular stationary
axially-symmetric solutions which admit finite energy density and quantum stability.
In the case of $SU(3)$ QCD the ansatz for regular axially-symmetric solutions
simplifies crucially the equations of motion and leads to a Weyl symmetric Lagrangian 
corresponding to a generalized $\phi^4$ model. 
A special subclass of Abelian stationary solutions with finite energy density 
is considered and has been proved to be stable against quantum gluon fluctuations.
  
\section{Axially-symmetric ansatz}

We consider a pure $SU(3)$ QCD Lagrangian and corresponding 
equations of motion
\bea
&&  {\cal L}_0 = -\dfrac{1}{4} F_{\mu\nu}^a F^{a\mu\nu}, \nn \\
&&  (D^\mu F_{\mu\nu})^a =0, \nn \\
&&  F_{\mu\nu}^a=\pro_\mu A_\nu^a-\pro_\nu A_\mu^a+g f^{abc} A_\mu^b A_\nu^c, \label{eqs0}
\eea
where $A_\mu^a$ ia a gauge potential, 
$(a,b,c=1,2,3)$ are color indices, and $\mu,\nu=1,2,3,4$ 
denote the space-time coordinates. 
One can generalize a known $SU(2)$
static axially symmetric Dashen-Hasslacher-Neveu (DHN) ansatz \cite{DHN} 
to the case of time-dependent solutions of $SU(3)$ Yang-Mills theory as follows
\bea
A_1^2&=&K_1,~~A_2^2=K_2,~~A_3^1=K_4,~~A_4^2=K_5, \nn \\
A_1^5&=&Q_1,~~A_2^5=Q_2,~~A_3^4=Q_4,~~A_4^5=Q_5, \nn \\
A_1^7&=&S_1,~~~A_2^7=S_2,~~A_3^6=S_4,~~~A_4^7=S_5, \nn \\
A_3^3&=&K_3,~~~A_3^8=K_8,   \label{SU3DHN}
\eea
where the three sets of off-diagonal components of the gauge potential
 $K_i,Q_i,S_i$ $(i=1,2,4,5)$ with Abelian gluon fields $K_{3,8}$ 
 correspond to $I,U,V$-type $SU(2)$ subgroups of $SU(3)$.
All fields  $K_i,Q_i,S_i$ are axially symmetric functions depending on
three coordinates $(r,\theta, t)$ (we use the standard spherical coordinates
($r,\theta, \varphi$). 
In the case of $SU(2)$ Yang-Mills theory
the DHN ansatz leads to equations of motion which are degenerate
due to the presence of a residual $U(1)$ gauge symmetry. We add a Lorenz type 
gauge fixing term to the original Lagrangian $ {\cal L}_0$ to fix the
appearance of such a residual symmetry after applying our ansatz 
\bea
{\cal L}_{gen}&=& {\cal L}_0 - \sum_{a=2,5,7} \dfrac{\alpha}{2} (\pro_r A^a_1
+\dfrac{1}{r^2} \pro_\theta A^a_2 -\pro_t A_4^a)^2,\nn \\
&& \label{Lagrgen}
\eea
where $\alpha$ is a gauge fixing number parameter. 
One can verify that the ansatz (\ref{SU3DHN}) is consistent with the Euler-Lagrange
equations obtained from the Lagrangian ${\cal L}_{gen}$
and leads to fourteen non-vanishing partial differential equations 
for  $K_i,Q_i,S_i$. It is suitable to set $\alpha=1$, in that case 
the linearized parts of the equations for  $K_i,Q_i,S_i$
contain the classical D' Alembert operator. 

It is surprising, that one can simplify further the obtained system of 
fourteen equations for the fields  $K_i,Q_i,S_i$ by applying the following
reduction ansatz
 \bea
& Q_{1,2,5}=-S_{1,2,5}=-K_{1,2,5},  ~~~~~~&K_{3,8}= -\frac{\sqrt 3}{2} K_4, \nn \\
& Q_{4}= \big (-\frac{1}{2}+\frac{\sqrt 3}{2}\big ) K_4, ~~~~~~~~~~~~~ &S_{4}= \big (-\frac{1}{2}-\frac{\sqrt 3}{2}\big ) K_4,  \label{reduction1}
 \eea
Substitution of this ansatz into all fourteen equations
for $K_i,Q_i,S_i$ results in four second order hyperbolic 
differential equations for four functions $K_{1,2,4,5}$
and one quadratic constraint containing first order derivatives
\bea
&&r^2 \pro_t^2 K_1-r^2 \pro_r^2 K_1-\pro_\theta^2 K_1+2r(\pro_t K_5-\pro_r K_1)+\cot\theta(\pro_r K_2-\pro_\theta K_1)+\dfrac{9}{2}\csc^2\theta K_4^2 K_1=0,  \label{eq1}
 \\
&&r^2 \pro_t^2 K_2-r^2 \pro_r^2 K_2-\pro_\theta^2 K_2+r^2 \cot\theta(\pro_t K_5-\pro_r K_1)-\cot\theta \pro_\theta K_2+\dfrac{9}{2}\csc^2\theta K_4^2 K_2=0, \label{eq2}
 \\
&&r^2 \pro_t^2 K_4-r^2 \pro_r^2 K_4-\pro_\theta^2 K_4+\cot\theta\pro_\theta K_4 
 +3r^2(K_1^2-K_5^2)K_4+3K_2^2K_4=0,\label{eq4} 
 \\
&&r^2 \pro_t^2 K_5-r^2 \pro_r^2 K_5-\pro_\theta^2 K_5+2r(\pro_t K_1-\pro_r  K_5)
+\cot\theta (\pro_t K_2-\pro_\theta K_5)+ \dfrac{9}{2}\csc^2\theta K_4^2 K_5=0, \label{eq5} 
\\
&&2 r^2 (K_5\pro_t K_4-K_1 \pro_r K_4 )+ K_2 (\cot \theta K_4-2 \pro_\theta K_4) 
+ K_4 (-\pro_\theta K_2 +r^2 (\pro_t K_5-\pro_r K_1))=0.  \label{constr1}
\label{eq5}
\eea
The ansatz is consistent with the original equations of motion of Yang-Mills theory.
Note that, if we substitute the ansatz into the original Lagrangian and then derive the Euler
equations for four independent fields $K_{1,2,4,5}$, certainly, we will not
obtain the constraint unless one introduces a Lagrange multiplier.  

To find a stationary solution one has to solve a boundary value problem
with unknown two-dimensional profile functions defining the boundary conditions.
Additional technical difficulties of numeric solving the above equations are caused by 
the non-linearity of the equations, the presence of the constraint  and slow 
numeric convergence of the solution in a three dimensional numeric domain. 
To overcome these obstacles we apply a method which allows to 
simplify the solving problem by transforming the equations on three-dimensional space-time
to equations on two-dimensional space. Such a method was applied in solving
equations for the sphaleron solution \cite{RR,KKB}.

First we use Fourier series representation for the functions 
$K_i(r,\theta,t),Q_i(r,\theta,t),S_i(r,\theta,t)$
 \bea
&&K_{i=1-4,8}(r,\theta,t)=\sum_{n=1,2,...} \tilde K_{i}^{(n)}(r,\theta)  \cos (n t), \nn \\
&&K_5(r,\theta,t)=\sum_{n=1,2,...} \tilde K_{5}^{(n)}(r,\theta)  \sin(nt),  \label{seriesdec}
\eea
and for $Q_i(r,\theta,t),S_i(r,\theta,t)$ one has similar decompositions.
Note that the series decompositions for $K_5,Q_5,S_5$ 
include only the basis functions $\sin(n t)$ due to the requirement of the energy density
to be finite and regular everywhere.
Substituting the series decompositions truncated at a finite order $n_f=N$ into the action with the
Lagrangian ${\cal L}_{gen}$ one can perform 
integration over the time period and polar angle, and obtain a reduced action 
\bea
S_{red}[\tK_i(r,\theta), \tQ_i(r,\theta), \tS_i(r,\theta)] =2 \pi \int dr d\theta \int_0^{2 \pi} dt {\cal L}_{gen}
\eea
 Taking variational derivatives of the reduced action with respect to the field modes
 $\tK_i^{(n)}, \tQ^{(n)}_i, \tS^{(n)}_i$ one can derive corresponding
 $4N$ Euler equations. 
A crucial advantage of our approach in solving the original equations motion 
is that one can impose an additional constraint on Fourier series
decompositions for the fields  $K_i,Q_i,S_i$ and simplify more
the structure of the reduced equations. Namely, we set all even Fourier modes
$\tK_i^{2 k},  \tQ_i^{2 k},  \tS_i^{2 k}$ to be vanished identically. 
Certainly, such a constraint reduces the space of possible axially symmetric solutions.
One should stress that this constraint is consistent with all $4N$ Euler equations obtained 
from the reduced action $S_{red}[\tK_i, \tQ_i, \tS_i] $.
Now one can apply the reduction ansatz (\ref{reduction1}) to the $4N$ Euler equations for the Fourier
modes
 \bea 
&& \tQ_{1,2,5}^{(n)}=-\tS_{1,2,5}^{(n)}=-\tK_{1,2,5}^{(n)}, ~~~~~~~~~\tK_{3,8}^{(n)}= -\frac{\sqrt 3}{2} \tK_4^{(n)}, \nn\\
 &&\tQ_{4}^{(n)}= \big (-\frac{1}{2}+\frac{\sqrt 3}{2}\big ) \tK_4^{(n)}, ~~~~~~~~~~~\tS_{4}^{(n)}= \big (-\frac{1}{2}-\frac{\sqrt 3}{2}\big ) \tK_4^{(n)},  \label{reduction2}
 \eea
 where ($n=1,3,5, ...N$).
 It is remarkable, that the reduction ansatz produces exactly $4N$ equations for $4N$ odd modes  $\tK_i^{(n)}, \tQ^{(n)}_i, \tS^{(n)}_i$ without generation of any additional constraints.
 In the leading order decomposition one has only four  partial differential equations 
 for the leading modes $\tK_{i=1,2,4,5}^{(1)}(r,\theta)$
 \bea
&& \Big( \pro^2_r+\dfrac{2}{r}\pro_r+\dfrac{1}{r^2}\pro_\theta^2+\dfrac{1}{r^2}\cot\theta\pro_\theta+M^2
  \Big )\tK_1^{(1)} -\dfrac{2}{r} M \tK_5^{(1)}
   -\dfrac{1}{r^2}\cot\theta \pro_r \tK_2^{(1)} 
  -\dfrac{27}{8r^2 \sin^2\theta} \tK_1^{(1)}(\tK_4^{(1)})^2 =0,\nn\\
&& \Big( \pro^2_r+\dfrac{1}{r^2}\pro_\theta^2+\dfrac{1}{r^2}\cot\theta\pro_\theta+M^2
  \Big )\tK_2^{(1)}   -M\cot \theta \tK_5^{(1)} 
      +\cot\theta \pro_r \tK_1^{(1)} 
  -\dfrac{27}{8r^2 \sin^2\theta} \tK_2^{(1)}(\tK_4^{(1)})^2 =0,\nn\\
&&  \Big( \pro^2_r+\dfrac{1}{r^2}\pro_\theta^2-\dfrac{1}{r^2}\cot\theta\pro_\theta+M^2
  \Big )\tK_4^{(1)} +\dfrac{3}{4}\tK_4^{(1)}(\tK_5^{(1)})^2 
       -\dfrac{9}{4r^2} \tK_4^{(1)}((r\tK_1^{(1)})^2+(\tK_2^{(1)})^2)
 =0,\nn\\
&&  \Big( \pro^2_r+\dfrac{2}{r}\pro_r+\dfrac{1}{r^2}\pro_\theta^2+\dfrac{1}{r^2}\cot\theta\pro_\theta+M^2
  \Big )\tK_5^{(1)} +\dfrac{2}{r} M \tK_1^{(1)} 
  +\dfrac{M}{r^2}\cot\theta \tK_2^{(1)}
   -\dfrac{9}{8r^2\sin^2\theta} \tK_5^{(1)}(\tK_4^{(1)})^2=0. \label{eqleading}
\eea

The system of equations  (\ref{eqleading}) admits a wide class of regular stationary solutions.
In particular, there is a class of regular solutions with a finite energy density and
different parities under the reflection symmetry $\theta \rightarrow \pi-\theta$
\bea
&&\tK_{1,5}(r,\pi-\theta) = \pm \tK_{1,5}(r,\theta) , \nn \\
&&\tK_2(r,\pi-\theta) =\mp \tK_4(r,\theta),\nn \\
&&\tK_4(r,\pi-\theta) =\tK_4(r,\theta),  \label{refl}
\eea
where the field mode $\tK_4(r,\theta)$ corresponding to the Abelian components $A_\mu^{3,8}$
of the gauge potential is invariant under the reflection transformation.
We call solutions corresponding to the lower and upper signs in (\ref{refl})  as type I and type II 
solutions respectively. An example of type I solution has been obtained in \cite{plb2017}, in the next subsection 
we describe type II solution.

\subsection{ Type II stationary solution} 

To solve numerically the equations (\ref{eqleading})  we choose
a rectangular numeric domain $(0\leq r\leq L, 0\leq \theta \leq \pi)$
and impose the following boundary conditions
 \bea
&&\tK_i^{(n)}(r,\theta)|_{r=0}=0, ~~~\tK_i^{(n)}(r,\theta)|_{\theta=0, \pi}=0. \label{bc1}
\eea
Solving the equations  (\ref{eqleading}) in the asymptotic region at far distance 
one can obtain asymptotic solution profiles of the functions $\tK_i^{(1)}$ 
\bea
&&\tK_1^{(1)}\simeq a_1^{(1)}(\theta) \dfrac{ \sin(r)}{r^2}, \nn \\
&&\tK_{2,4}^{(1)}\simeq a_{2,4}^{(1)} (\theta) \cos(r), \nn \\
&&\tK_5^{(1)}\simeq a_5^{(1)}(\theta)  \dfrac{ \cos(r)}{r^2}, \label{asym1}
\eea
where $a_i^{(1)}(\theta)$ are arbitrary periodic functions depending on the polar angle. 
It is clear, that there is a wide class
of regular solutions determined by the choice of the angle functions  $a_i^{(1)}(\theta)$.
We are interested in solutions with the lowest angle modes since such classical solutions   
correspond to the QCD vacuum.

We use the iterative Newton method which starts with some initial profile 
functions and after proper number of iterations produces a convergent numeric solution 
to a given boundary value problem for a set of elliptic partial differential equations.
In the initial profile functions for  $\tK_{i}^{(1)}$ we choose the lowest angle modes for 
$a_i^{(1)}(\theta)$  consistent with the finite energy density condition
\bea
&&a_{1,4,5}^{(1)}(\theta)=c_{1,4,5} \sin^2\theta, \nn \\
&&a_{2}^{(1)}(\theta)=c_2 \sin (2\theta)
\eea
An advantage of the iterative method is that the obtained numeric solution 
is not much sensitive to chosen initial profile functions, especially to the shape of the angle 
modes $a_i^{(1)}(\theta)$ and number values
of the integration constants $c_i$. Remind, that in the case of non-linear partial
differential  equations the regular solutions exist typically only for some special sets of integration 
constants.
With this we solve numerically the equations (\ref{eqleading}), 
the solution is presented in Figs. 1,2,3.
%
\begin{figure}[h!]
\centering
\subfigure[~]{\includegraphics[width=42mm,height=30mm]{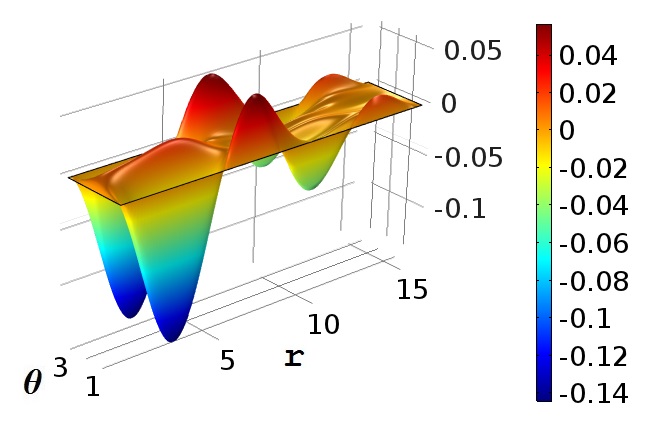}}
\subfigure[~]{\includegraphics[width=42mm,height=30mm]{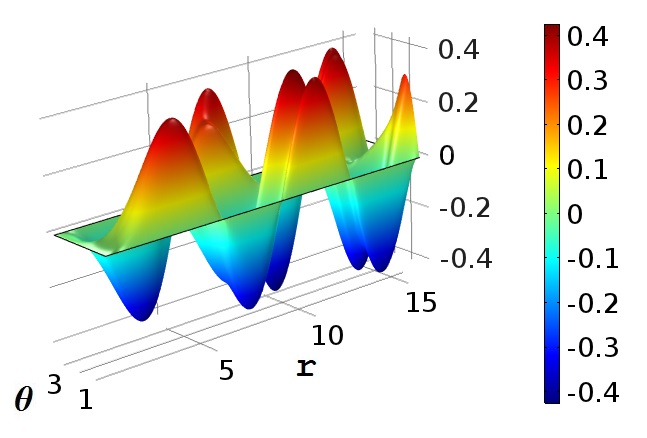}}
\subfigure[~]{\includegraphics[width=42mm,height=30mm]{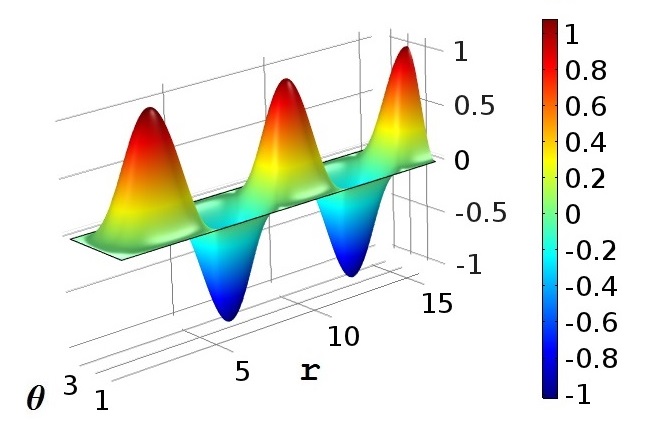}}
\subfigure[~]{\includegraphics[width=42mm,height=30mm]{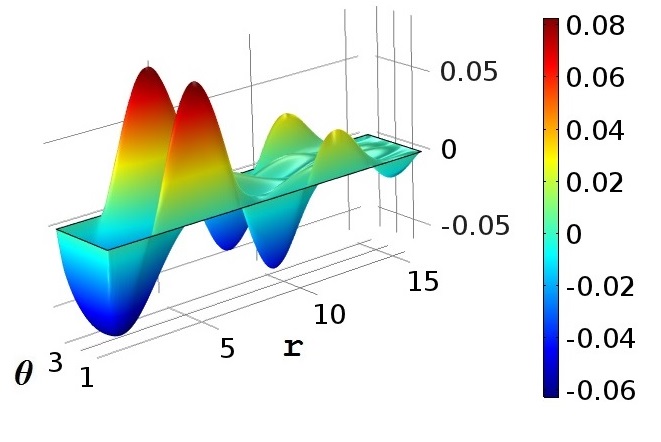}}
\caption[fig1]{Solution profile functions in the leading order: (a) $\tK_{1}^{(1)}$;
(b)  $\tK_{2}^{(1)}$; (c)  $\tK_{4}^{(1)}$; (d)   $\tK_{5}^{(1)}$ ($g=1, M=1$).}\label{Fig1}
\end{figure}
\begin{figure}[h!]
\centering
\subfigure[~]{\includegraphics[width=42mm,height=30mm]{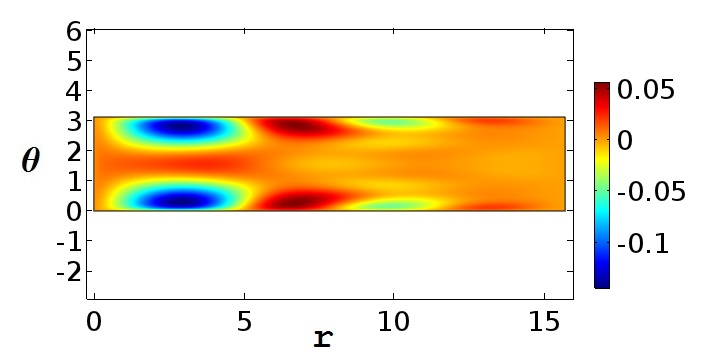}}
\subfigure[~]{\includegraphics[width=42mm,height=30mm]{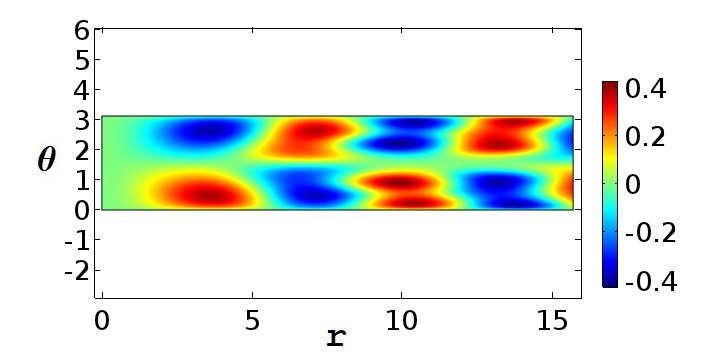}}
\subfigure[~]{\includegraphics[width=42mm,height=30mm]{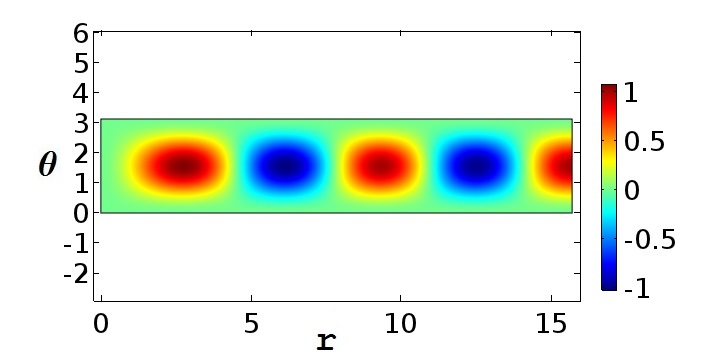}}
\subfigure[~]{\includegraphics[width=42mm,height=30mm]{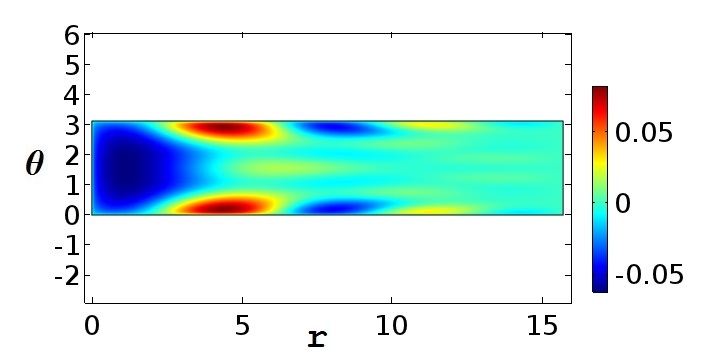}}
\caption[fig2]{Contour plots for the solution profile functions: (a) $\tK_{1}^{(1)}$;
(b)  $\tK_{2}^{(1)}$; (c)  $\tK_{4}^{(1)}$; (d)   $\tK_{5}^{(1)}$.}\label{Fig2}
\end{figure}
\begin{figure}[h!]
\centering
\subfigure[~]{\includegraphics[width=65mm,height=40mm]{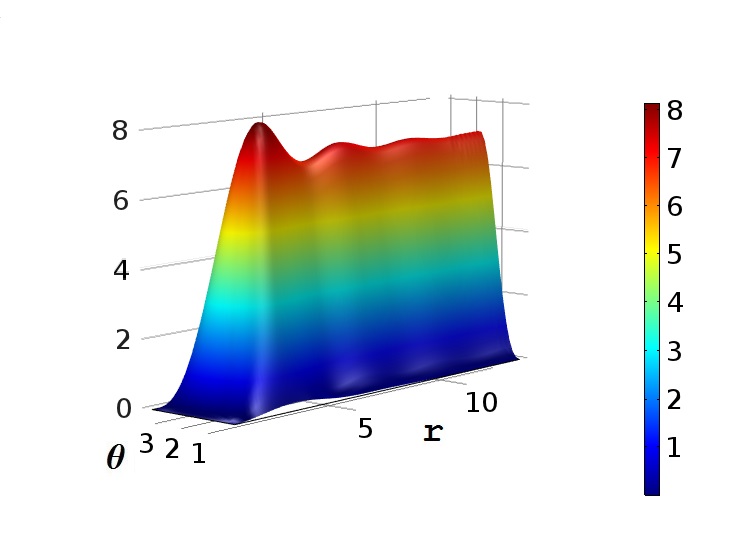}}
\subfigure[~]{\includegraphics[width=52mm,height=40mm]{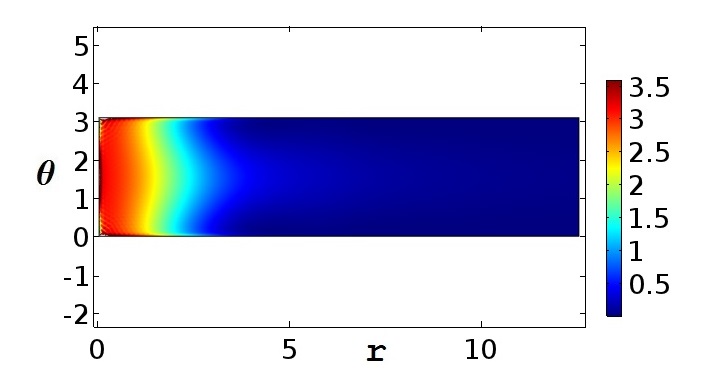}}
\caption[fig3]{(a)  An integral energy density $r^2 \sin\theta{\cal E}$ (averaged over the time period); 
(b) a contour plot for the time averaged energy density ${\cal E}$ ($g=1, M=1$).}\label{Fig3}
\end{figure}

 The energy density is decreasing along the radial direction as $\dfrac{1}{r}$, and it has a maximum 
 at the origin, and the total energy grows up linearly with increasing the
 radial size of a chosen numeric domain. Integration over the numeric domain constrained by $L=4 \pi$
 produces a value of the total energy $E_{II}^{tot}=121.6$ (up to multiplier factor $2 \pi$ due to further integration over the azimuthal angle $\varphi$). 
 Our numeric analysis of the solutions implies that solutions are determined by two parameters:
 the conformal parameter $M$ and the asymptotic amplitude $A_0$ of the Abelian field
 component mode $K_4(r,\theta, t)$. We fix the values of $M$ and $A_0$ to one, the amplitudes 
 for other fields $K_{1,2,5}$ are obtained from the numeric solution. This allows to compare
 solutions with fixed values of $M,A_0$ and with different parities by evaluating their energies. 
 The energy density of the stationary monopole-antimonopole pair solution with an opposite parity proposed in \cite{plb2017} has nearly the same shape and a total energy $E_{I}^{tot}=121.9$ 
 which is very close to the value of $E_{II}^{tot}$.
Note that a dominant contribution to the energy is 
provided by the field mode $K_4(r,\theta,t)$, it is $E(K_4)=117.67$ (in units $g=1, M=1$).
One should stress that our solution is completely different
 from known non-linear standing wave type solutions which have singularities.

  \subsection{Abelian regular stationary axially-symmetric solutions}
  
  Solutions determined by the boundary conditions (\ref{bc1}) 
  correspond to regular single-valued functions. Since the fields 
  $K_i(r,\theta,t)$ represent components of the gauge potential
  which are not physical observables (unless the color symmetry is broken), one can choose
  boundary conditions with multi-valued intial profile functions as well.
The gauge invariant quantities (like the energy, action etc) must be regular everywhere.  
In the leading order approximation one can find 
  a local solution to the equations   (\ref{eqleading}) 
  near the origin $r=0$ in terms of the Taylor series expansion
  \bea
&&\tK_1^{(1)}=c_1 (\dfrac{\pi}{2}-\theta)+\dfrac{1}{2}r^2 (-c_1(\dfrac{\pi}{2}-\theta)+c_5 \cos \theta)+O(r^3), \nn \\
&&\tK_2^{(1)}=-c_1 r+O(r^3), \nn \\
&&\tK_4^{(1)}=c_4 r^2 \sin^2 \theta +O(r^3), \nn \\
&&\tK_5^{(1)}=-c_1 r (\dfrac{\pi}{2}-\theta)+c_5 \cos \theta)+O(r^3),   \label{bc2}
\eea
where $c_{1,4,5}$ are arbitrary integration constants. One can verify that 
the local solution provides a regular energy density near the origin 
 We impose periodic boundary conditions along the boundaries $(\theta = 0, \pi)$
 and the same asymptotic conditions (\ref{asym1}).
 Wth this one can solve the system of equations  (\ref{eqleading}),
the obtained solution is presented in Fig. 4. 
 \begin{figure}[h!]
\centering
\subfigure[~]{\includegraphics[width=40mm,height=30mm]{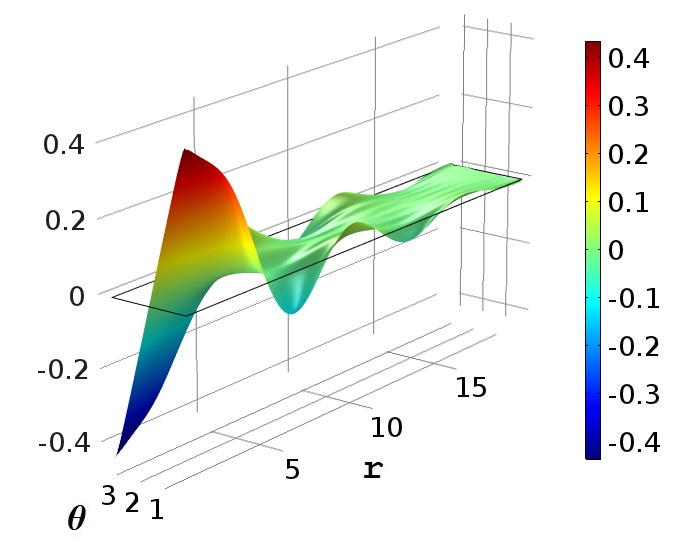}}
\hfill
\subfigure[~]{\includegraphics[width=40mm,height=30mm]{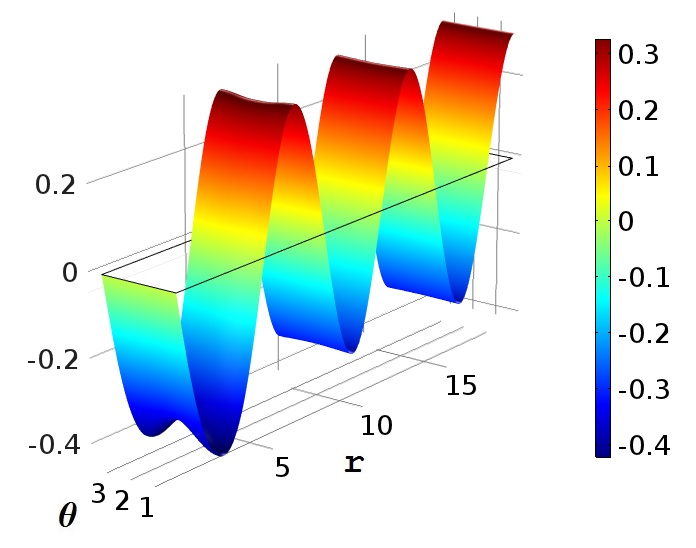}}
\hfill
\subfigure[~]{\includegraphics[width=40mm,height=30mm]{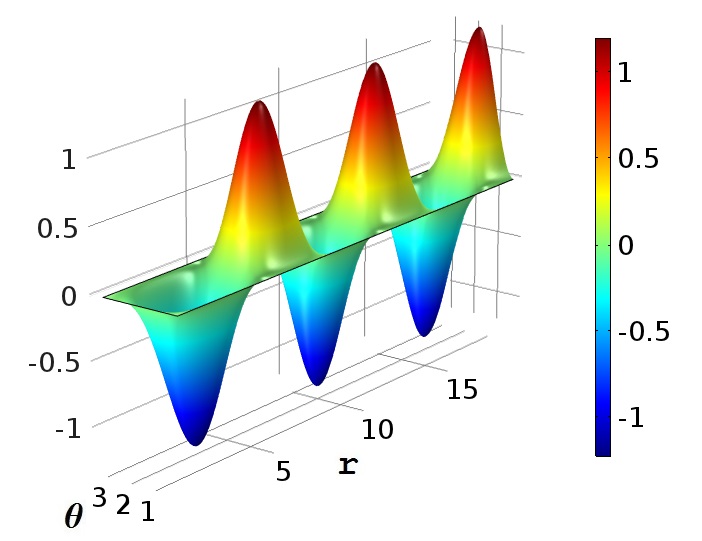}}
\hfill
\subfigure[~]{\includegraphics[width=40mm,height=30mm]{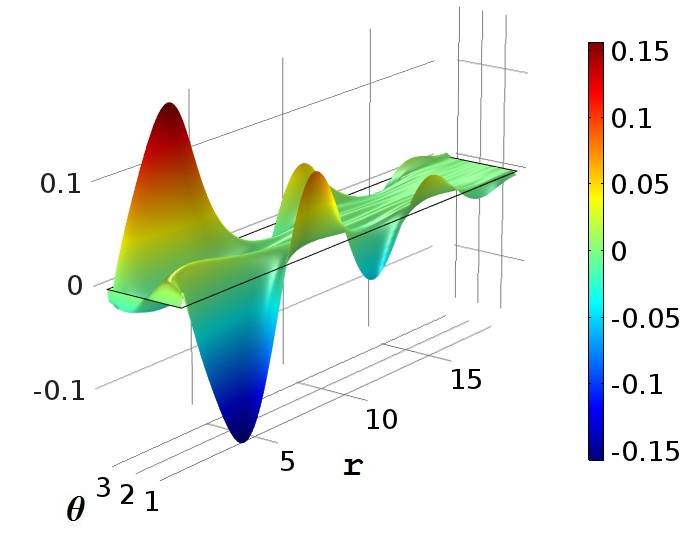}}
\hfill
\caption[fig4]{Solution in the leading order: (a) $K_{1}^{(1)}$;
(b)  $K_{2}^{(1)}$; (c)  $K_{4}^{(1)}$; (d)   $K_{5}^{(1)}$.}\label{Fig4}
\end{figure}
The solution implies that azimuthal field strength components represent multi-valued functions
along the $Z$-axis. However, a corresponding energy density
is regular everywhere,  and it has a similar shape as one in Fig. 3.
A class of such solutions is determined by values of three parameters characterizing the
asymptotic behavior, namely, by the conformal parameter $M$ and asymptotic 
amplitudes $a_{04}, a_{02}$ of the oscillating modes $K_{2,4}$. 
Contrary to the case of type II solution, Fig. 1-3, the asymptotic amplitude $a_{02}$
of the mode $\tK_2^{(1)}$  is an additional free number parameter. In the limit $a_{02}\rightarrow 0$
the modes  $\tK_{1,2,5}^{(1)}$ vanish identically, and one results in a solution which satisfies 
an Abelian type partial differential equation
 \bea
&&\pro_t^2 K_4-\pro_r^2 K_4-\dfrac{1}{r^2}\pro_\theta^2 K_4+\dfrac{\cot\theta}{r^2}\pro_\theta K_4 =0.\label{eqMaxwell} 
\eea
In the case of $SU(2)$ Yang-Mills theory the equation (\ref{eqMaxwell})
represents an equation of motion  for one non-vanishing gauge field component $A_\varphi^3$.
So, the Eq. (\ref{eqMaxwell}) coincides identically with the equation of motion
for one non-zero vector potential $A_\varphi$ of the Maxwell theory.

Let us consider in a detail the Abelian type solutions to Eq. (\ref{eqMaxwell}). 
It is clear that solutions to this equation defines a corresponding class of non-Abelian
solutions to the full set of $SU(2)$ or $SU(3)$ equations of motion within the reduction ansatz (\ref{SU3DHN},\ref{reduction1}). 
A basis in the vector space of regular solutions to Eq. (\ref{eqMaxwell})
is formed by the following functions:
\bea
K_{4}^{(k)}(r,\theta,t)&=&R_k(r)T_k(\theta) \sin(t), \nn \\
R_k(r) &=&\sqrt {r} J_{\frac{2k+1}{2}}(r), \nn \\
T_k(\theta)&=&\left \{
\ba{lcll}
 _2F^1[-\dfrac{k+1}{2}, \dfrac{k}{2};\dfrac{1}{2};\cos^2\theta],    &\rm{for~ odd~} k,&  \\
\cos\theta~_2F^1[-\dfrac{k}{2}, \dfrac{k+1}{2};\dfrac{3}{2};\cos^2\theta] ,  &\rm{for~ even~} k,&
\ea
\right .
\eea
where $J_n(r)$ is the Bessel function of the first kind, $_2F_1(a,b;c;z)$ is the hypergeometric function,
 $k=1,2,3,...$, and integration constants $C_{1,2}$ are chosen in such a way
to provide regular field configurations. 
One can write down first three basis solutions corresponding to values $k=1,2,3$
\bea
K_4^{(1)}(r,\theta,t)&=&(-\cos r+\dfrac{1}{r}\sin r) \sin^2\theta \sin t, \nn \\
K_4^{(2)}(r,\theta,t)&=&  \dfrac{1}{r^2}\big (3 r \cos r+(-3+r^2) \sin r \big )   \cos\theta \sin^2\theta \sin t, \nn \\
K_4^{(3)}(r,\theta,t)&=&\dfrac{1}{r^3}\big (r(r^2-15)\cos r+3 (5-2r^2)\sin r \big) \big (3+5 \cos (2 \theta)\big) \sin^2\theta \sin t.
\label{Abelsol}
\eea
The lowest mode $K_4^{(1)}(r,\theta,t)$ provides an interpolating function with a high accuracy for the 
numeric solution presented in the previous subsection, Fig. 1c. One can calculate 
the contribution of that mode to the total energy density in the numeric domain restricted
by the parameter value $L=4 \pi$. The total energy has a value $E(K_4^{(1)})=117.685$ which is 
very close to the value $E(K_4)=117.67$ obtained from the numeric solution. The difference between two functions
is 1.08 \% by norm.
Note that one has conformal classes of regular sttaionary 
solutions generated by the scaling transformation $r\rightarrow M r, t\rightarrow M t$
of the solutions (\ref{Abelsol}).

  
\section{Weyl symmetric structure of the reduced Lagrangian}

Let us consider first a simple case of a pure $SU(2)$ QCD.
The corresponding Lagrangian can be written in explicit Weyl symmetric form
using complex notations for the off-diagonal components of the gauge potential
\bea
{\cal L}&=&-\dfrac{1}{4} F_{\mu\nu}^2-\dfrac{1}{2}|{D}_\mu{W}_\nu-{D}_\nu{W}_\mu|^2  
+ig F_{\mu\nu} \astW_\mu W_\nu-\dfrac{g^2}{2} \Big[(\astW_\mu W_\mu)^2 
 -(\astW_\mu)^2 (W_\nu)^2 \Big], 
\label{SU2Lagr}
\end{eqnarray}
where
\bea
&& W_{\mu\nu} = - i g ( \astW_\mu W_\nu - \astW_\nu W_\mu ), \nn\\
&&{D}_\mu{W}_\nu = (\partial_\mu + ig A^3_\mu) W_\nu, \nn \\
&& W_\mu= \dfrac{1}{\sqrt 2} (A^1_\mu+i A_\mu^2).
\eea
The Weyl symmetry is represented by the reflection 
transformation of the Abelian potential, $A_\mu^3\rightarrow -A_\mu^3$.
A generalized DHN ansatz reads 
\bea
A_1^2&=&K_1,~~A_2^2=K_2,~~A_3^3=K_3,~~~A_3^1=K_4,~~A_4^2=K_5.    \label{SU2DHN}
\eea
After substitution of this ansatz into the Lagrangian
one has cubic interaction terms
\bea
{\cal L}_{cubic}=\dfrac{1}{r^2 \sin^2 \theta} \Big [K_1 (K_4 \pro_r K_3-K_3 \pro_r K_4)+
K_2 (K_4 \pro_\theta K_3-K_3 \pro_\theta K_4)+
K_5 (-K_4 \pro_t K_3-K_3 \pro_t K_4) \Big ].
\eea
The presence of the cubic interaction terms
leads to breaking of the Weyl symmetry.
As a consequence, the stationary monopole pair solution
in a pure $SU(2)$ QCD does not possess Weyl symmetry. 
However, since the field component $K_4$
is suppressed  in the leading order approximation, the Weyl symmetry
of the Lagrangian takes place approximately. 

Let us consider the case of the $SU(3)$ QCD. 
The Lagrangian can be written in complex notations 
as follows
\bea
{\cal L}_0&=&\sum_{p=1,2,3}\Big \{  -\dfrac{1}{6} (F_{\mu\nu}^p)^2-\dfrac{1}{2}| D^p_\mu W_\nu^p-
D_\nu^p W_\mu^p|^2  -ig F_{\mu\nu}^p W_\mu^{*p} W_\nu^p\Big \}+{\cal L}_{int}^{(4)}[W] , \nn \\
F_{\mu\nu}^p& =& \pro_\mu B_\nu^p-
\pro_\nu B_\mu^p, ~~~~~~D^p_{\mu}W_\nu^p =
(\pro_\mu - ig B_\mu^p) W_\nu^p,  ~~~~~~
B_\mu^p=A_\mu^\alpha \rt_{\alpha}^{\,p}, \nn \\
W^{1}_{\mu}&=& \dfrac{1}{\sqrt 2} (A_\mu^1 + i A_\mu^2), ~~~~~
W^{2}_{\mu}= \dfrac{1}{\sqrt 2} (A_\mu^6 + i A_\mu^7), ~~~~~
W^{3}_{\mu}= \dfrac{1}{\sqrt 2} (A_\mu^4 - i A_\mu^5), \label{LWW}
\eea
where the index $(p=1,2,3)$ corresponds to linear combinations of the gauge potentials 
which form the representation of the Weyl permutation group, and
$\rt_\alpha^p$ are the root vectors of the Lie algebra of $SU(3)$,
the index $\alpha$ takes two values,
($\alpha=3,8$), corresponding to the generators of the Cartan  algebra of $SU(3)$ 
\bea
&&\rt_{\alpha}^{\,1} = (1,0), ~~~~\rt_{\alpha}^{\,2} =
(-1/2,\sqrt 3/2), ~~~~ \rt_{\alpha}^{\,3} = (-1/2,-\sqrt 3/2) .
\eea

First of all, note that the Lagrangian of a pure $SU(3)$ QCD in the form (\ref{LWW}) can not be written
in the Weyl symmetric form since the interaction term ${\cal L}_{int}^{(4)} $
is not factorized into a direct sum of parts corresponding to separated Weyl sectors. 
It is remarkable that applying the ansatz (\ref{reduction1}) one obtains an explicit Weyl symmetric
reduced Lagrangian ${\cal L}_{red}(K_{1,2,4,5})$. In particular, the quartic interaction term originating from the kinetic part
in (\ref{LWW}) coincides identically with the Weyl symmetric expression for ${\cal L}_{int}^{(4)} $ 
\bea
{\cal L}_{int}^{(4)}[W]=-\dfrac{9}{8}\sum_{p=1,2,3} \Big [
 (W^{*p\mu} W^p_\mu)^2-(W^{*p\mu} W^{*p}_\mu) (W^{p\nu} W^p_\nu) 
\Big ]. 
\eea
Another essential feature of the reduced Lagrangian  ${\cal L}_{red}(K)$ is that
all cubic interaction terms are mutually canceled, in particular, the third term in   (\ref{LWW})
vanishes identically itself. As it is known,  such a term represents an anomalous magnetic moment interaction
which is responsible for the instability of the Savvidy vacuum \cite{savv,N-O}. Vanishing of this term gives
an additional indication that stationary monopole-like solution could be stable under the gluon vacuum fluctuations.
Indeed, it has been proved recently, that the vacuum made of stationary monopole-antimonopole pair 
is stable \cite{plb2017}.  
With this, the final expression for the
reduced Lagrangian takes the following form
 \bea
{\cal L}_{red}&=&\sum_{p=1,2,3}\Big \{  -\dfrac{1}{6} (F_{\mu\nu}^p)^2-\dfrac{1}{2}| \pro_\mu W_\nu^p-
\pro_\nu^p W_\mu^p|^2 -\dfrac{9}{4}\sum_{p=1,2,3} \Big [
 (W^{*p\mu} W^p_\mu)^2-(W^{*p\mu} W^{*p}_\mu) (W^{p\nu} W^p_\nu) \Big ].  \label{Lphi4}
 \eea
 The Lagrangian can be written in terms of four real independent fields $K_i(r,\theta,t)$
 \bea
 {\cal L}_{red}(K)&=&  \dfrac{3}{2r^2} \Big [ r^2 (\pro_t K_1 -\pro_r K_5)^2 -\pro_\theta K_1^2+\pro_\theta K_5^2 \Big ]
+  \dfrac{3}{2 r^2} \Big [ \pro_t K_2 (\pro_t K_2-2 \pro_\theta K_5) - \pro_r K_2(\pro_r K_2-2 \pro_\theta K_1) \Big ]\nn \\
     &+&  \dfrac{9}{4r^4 \sin^2 \theta}\Big [ r^2 (\pro_t K_4^2-\pro_r K_4^2) +\pro_\theta K_4^2 \Big] -
        \dfrac{27}{4 r^4\sin^2 \theta} \Big [K_4^2 (K_2^2+r^2 (K_1^2-K_5^2) \Big ].
 \eea
 One can observe immediately that the Lagrangian $ {\cal L}_{red}(K)$ belongs to a field model
with a simple quartic potential without derivatives. So that, on the space of special stationary solutions
one has embedding of $\phi^4$ type model into the Yang-Mills theory. 

\section{Quantum stability}

For simplicity we consider the quantum stability of the stationary wave type solution 
$K_4^{(1)}(r,\theta,t)$, (\ref{Abelsol}), under small quantum gluon fluctuations
in the case of a pure $SU(2)$ QCD. 
A general $SU(2)$ gauge potential $A_\mu^a$ is split into a sum of
a classical background field $B_\mu^a$ and fluctuating quantum part $Q_\mu^a$.
The background field represents the stationary solution
\bea
B_\mu^a=\delta_{\mu 3} \delta^{a,3} K_4^{(1)}(r,\theta,t).
\eea
The ``Schr\"{o}dinger'' type equation for possible unstable quantum modes is the following
\cite{plb2017} 
\bea
\Ka_{\mu}^{ab} \Psi_\nu^b=\lambda \Psi_\mu^a, \label{schr3}
\eea
where the operator $\Ka_{\mu}^{ab} $ corresponds to one-loop gluon contribution 
to the effective action \cite{plb2017}
\bea
\Ka_{\mu\nu}^{ab}&=&-
\delta^{ab} \delta_{\mu\nu} \pro^2_t
            -\delta_{\mu\nu}({\Da}_\rho {\Da}^\rho)^{ab} -2 f^{acb}{\mathcal
	    F}_{\mu\nu}^c, \quad \label{Koper}
\eea
where the covariant derivative $\Da_\mu$ and field strength $\Fa_{\mu\nu}$ 
are defined in terms of the classical background solution.  
The existence of solutions to Eq. (\ref{schr3}) with negative eigenvalues $\lambda$
would indicate to the presence of unstable modes which destabilize the classical solution. 

We choose a temporal gauge for the quantum gauge potential, this simplifies the matrix part of the
operator $K_{\mu\nu}^{ab}$ and reduces the number of equations to nine elliptic
second order partial differential equations on the three-dimensional domain 
$(0\leq r \leq L, 0\leq \theta \leq \pi, 0\leq t \leq \dfrac{2\pi}{M})$.
Direct substitution of the classical solution 
\bea
K_4^{(1)}(r,\theta,t)&=&A_0(-\cos (M r)+\dfrac{1}{M r}\sin (M r)) \sin^2\theta \sin (M t)\equiv f_0(r,\theta,t),
\eea
into the eigenvalue equation (\ref{schr3}) leads to factorization of the initial nine equations to
three independent sets of equations which include the following functions: (I) $\Psi_1^1,\Psi_2^1,\Psi_3^2$,
(II) $\Psi_1^2,\Psi_2^2,\Psi_3^1$, (III) $\Psi_m^3$ (m=1,2,3). The last group of equations
corresponding to the Abelian direction in the color space represents free equations and do not
produce negative modes. The second set of equations becomes identical to the first
set of equations after changing variables $\Psi_1^2\rightarrow \Psi_1^1, \Psi_2^2 \rightarrow \Psi_2^1,
\Psi_3^1 \rightarrow \Psi_3^2$ and reflection of the background field, $f_0\rightarrow - f_0$.
So one has to solve only one system of three eigenvalue equations 
\bea
&&-\Delta \Psi_1^1 +\dfrac{1}{r^2} \Big ((2+\csc^2 \theta f_0^2)\Psi_1^1+2(\cot\theta+ \pro_\theta )\Psi_2^1-
2 \csc \theta (f_0-r\pro_r f_0)\Psi_3^2 \Big )=\lambda \Psi_1^1, \nn \\
&&-\Delta \Psi_2^1 +\dfrac{1}{r^2} \Big (\csc^2\theta(1+f_0^2) \Psi_2^2-2 \pro_\theta \Psi_1^1
-2\csc\theta(\cot \theta f_0-\pro_\theta f_0) \Psi_3^2\Big )=\lambda \Psi_2^1, \nn \\
&&-\Delta \Psi_3^2 +\dfrac{1}{r^2} \Big ((1+\cot^2\theta+\csc^2\theta f_0^2)\Psi_3^2
-2\csc \theta (f_0   - r\pro_r f_0) \Psi_1^1
-2\csc \theta (\cot\theta  f_0-\pro_\theta f_0) \Psi_2^1 \Big )=\lambda \Psi_3^2, \nn \\
&& \Delta \equiv \pro^2_t+\pro^2_r +\dfrac{2}{r}\pro_r +\dfrac{1}{r^2} \pro^2_\theta+\dfrac{\cot \theta}{r^2} \pro_\theta,
\label{3Psi}
\eea
where $\Delta$ is a part of the vector Laplace operator.
The system of equations (\ref{3Psi}) corresponds to a quantum mechanical potential problem 
of three interacting particles. The equations contain positive centrifugal potentials depending on 
space coordinates $(r,\theta)$ and two different attractive potentials 
\bea
V_1&=& -\dfrac{2\csc \theta}{r^2} (f_0-r \pro_r f_0), \nn \\
V_2&=&-\dfrac{2\csc \theta}{r^2}(\cot\theta f_0-\pro_\theta f_0).
\eea
One can verify that potentials $V_1$ and $V_2$ have no dangerous singularities at the origin $r=0$,
they are finite everywhere and decrease along the radial direction as $\dfrac{1}{r}$ and $\dfrac{1}{r^2}$ 
respectively. It is clear, that such potentials lead to a positive eigenvalue spectrum 
at small enough values of the parameters $A_0, M$.
Exact numeric solving the system of equations confirms absence of negative modes, Fig. 5.
\begin{figure}[h!]
\centering
\subfigure[~]{\includegraphics[width=50mm,height=38mm]{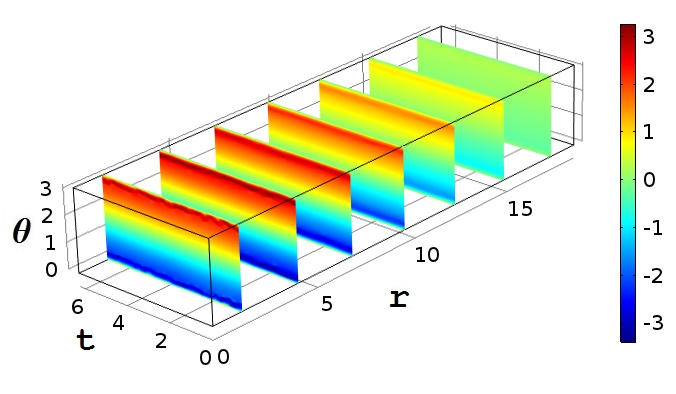}}
\hfill
\subfigure[~]{\includegraphics[width=50mm,height=38mm]{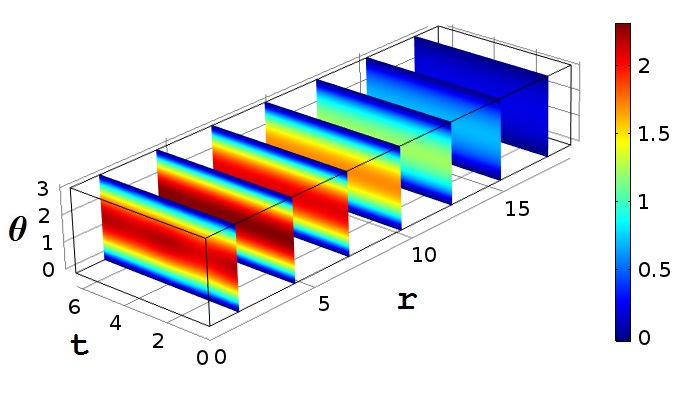}}
\hfill
\subfigure[~]{\includegraphics[width=50mm,height=38mm]{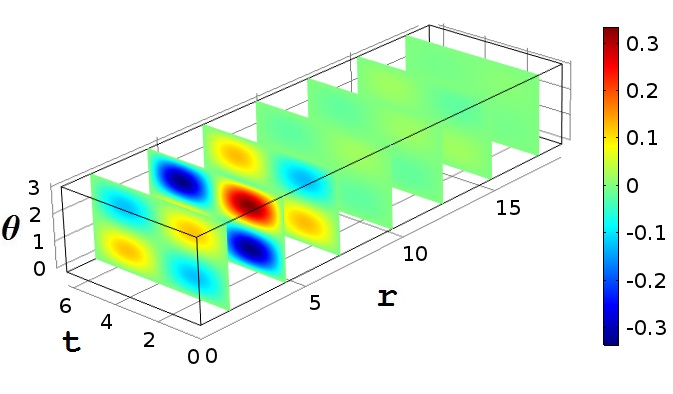}}
\hfill
\caption[fig5]{Eigenfunctions corresponding to the lowest
eigenvalue $\lambda=0.0298$: (a) $\Psi_1^1$; 
 (b) $\Psi_2^1$; (c) $\Psi_3^2$.
}\label{Fig5}
\end{figure}

In conclusion, we propose a new class of regular stationary solutions with a finite energy density 
in a pure $SU(3)$ QCD. Recently it has been proved that the stationary spherically symmetric 
monopole and monopole-antimonopole pair solutions are 
stable against small quantum gluon fluctuations \cite{plb2017, prd2017}. We expect that the whole 
class of considered regular stationary solutions possesses quantum stability as well. 
We have considered a class of regular Abelian 
stationary solutions and have proved thier stability under small quantum gluon fluctuations.
Since the Abelian solutions possess the classical stability as well, they provide
the most preferable field configurations for the QCD vacuum in quasiclassical approximation. 
We suppose that the regular stationary solutions play an important role in microscopic 
description of the QCD vacuum formation. This issue will be considered in the forthcoming paper.

\acknowledgments
 
 One of authors (DGP) thanks Prof. C.M. Bai
for warm hospitality during his staying in Chern Institute of Mathematics
 and Dr. Ed. Tsoy for useful discussions of numeric aspects.
 The work is supported by the grant OT-$\Phi$2-10.

\end{document}